\documentclass[cameraready]{Interspeech}

\title{In Defense of Using Worst-case Privacy Disclosure as Privacy Evaluation Metric of Voice Anonymization}

\author[affiliation={1}, orcid=0000-0001-8246-0606]{Xin}{Wang}
\author[affiliation={2}, orcid=0000-0002-6645-6524]{Xiaoxiao}{Miao}

\address{
    $^1$ National Institute of Informatics, Japan \\
    $^2$ Duke Kunshan University, China
}

\email{wangxin@nii.ac.jp}


\usepackage{comment}
\usepackage{multirow}
\usepackage{tikz}
\usetikzlibrary{shapes.geometric, arrows.meta, positioning, calc}

\newcommand{\costmiss}{C_{\text{fr}}^{\classtar{}}}
\newcommand{\costfanon}{C_{\text{fa}}^{\classnon{}}}

\newcommand{\symbolhp}{{H}}

\newcommand{\classtar}{\mathsf{Y}}
\newcommand{\classnon}{\mathsf{N}}

\newcommand{\evidence}{\boldsymbol{x}}
\newcommand{\score}{s}

\newcommand{\acty}{\mathtt{AFFIRM}}
\newcommand{\actn}{\mathtt{DENY}}

\begin{document}

\maketitle

\begin{abstract}

The voice anonymization community mainly uses Equal Error Rate (EER) to evaluate the performance of voice identity protection. While alternative metrics such as privacy-ZEBRA and a rank-based metric have been proposed, their underlying assumptions and differences may not be well known, especially to newcomers. This paper is motivated to fill the gap. 
Based on the concept of Shannon's perfect secrecy (or privacy), this paper positions itself as a defense of the privacy-ZEBRA framework. While no new metric is proposed, 
this paper explains how an `ideal' system in terms of EER may fail to gauge the information leakage on individual speakers in the log-likelihood ratio (LLR) space. The paper also shows how the rank-based metric can be cast into a metric that follows the same principle of perfect secrecy and how their best solutions are equivalent. Furthermore, the paper explains how the method of estimating LLRs may affect the evaluation results. These discussions are, to the best of the authors' knowledge, not explored or explained in detail in existing papers. Last but not least, the findings are demonstrated on simulated and VoicePrivacy Challenge data.
\end{abstract}

\section{Introduction}
Speech signals convey varied types of information, including personally identifiable information (PII) about the speaker~\cite{NAUTSCH2019441,backstromPrivacy2025}. In response to the growing legislative demands (e.g., GDPR~\cite{Nautsch2019}) for protecting the PII, the speech research community has put great efforts in producing related technologies, most of which are referred to as voice (pseudo-)anonymization~\cite{tomashenko2020introducing,panarielloVoicePrivacy2024,tomashenkoThird2026a} or de-identification~\cite{jin2009speaker,yuan_deid-vc_2022, tavi2022improving,chen_voicecloak_2023}. 

Whatever the technical solution for protecting PII is, we need an evaluation metric to gauge how well the PII is kept in `secret', or, in other words, being free from unauthorized access. 
Historically, metrics designed for automatic speaker verification (ASV), especially the Equal Error Rate (EER), have dominated most of the recent studies~\cite{tomashenko2020introducing,panarielloVoicePrivacy2024,tomashenkoThird2026a,yuan_deid-vc_2022,tavi2022improving}. A perfect-secrecy-based metric called privacy-ZEBRA~\cite{Nautsch2020} was proposed around the time of the first VoicePrivacy initiative, but it was not widely adopted thereafter. A more recent ranking-based metric~\cite{backstromPrivacy2026} was proposed to take into consideration the impact from other speakers in the dataset. There are also variants based on counting ASV errors~\cite{jin2009speaker,chen_voicecloak_2023,vauquierLegally2025, tsaprazlisVoxGuard2026a}.

A newcomer to the field is likely to ask what the most suitable metric is. Everyone may find their answer after reading the related papers, but many (including the authors of this paper) may feel puzzled due to different terminology, implicit assumptions, and required background knowledge across studies. 

This paper attempts to fill the gap.
It starts from a `first-principle' based on \emph{perfect secrecy} and discusses the aforementioned metrics in terms of their relationships and properties. 
After describing perfect secrecy and the threat model for voice privacy in \S~\ref{sec:secrecy}, the paper presents the following content:
\begin{itemize}
    \item \S~\ref{sec:zebra} recaps privacy-ZEBRA and its relationship to differential privacy (DP)~\cite{dwork2014algorithmic}.
    \item \S~\ref{sec:eer} is on how EER characterizes an aggregate metric of privacy protection but neglects the risk for individuals. The discussion is further divided into two parts, depending on whether \emph{scores} used to compute the EER are perfectly calibrated into the log-likelihood ratios (LLRs). A demonstration using simulated data is presented. Other metrics using fixed decision thresholds are briefly mentioned. 
    \item \S~\ref{sec:rank} discusses how a ranking-based metric can be cast into the perfect secrecy framework. It offers a new way of estimating LLRs based on the ranking. The equivalence between the ranking-based LLRs and those from privacy-ZEBRA is justified using simulated data. 
    \item \S~\ref{sec:calibration} discusses how the LLR estimation method affects the evaluation of privacy protection, particularly the non-linear score calibration method adopted by privacy-ZEBRA.
    \item \S~\ref{sec:vpc} shows results on VoicePrivacy 2024 systems~\cite{panarielloVoicePrivacy2024}.
\end{itemize}
{Note that much of the content before \S~\ref{sec:sub:eerllr} can find their roots in existing literature.} Readers familiar with Bayes decision and EER can simply skim through them. {However, a goal of this paper is to present them in a more accessible manner that bridges the gap between introductory and advanced contents scattered in different papers.} 
This is unique among existing literature in the humble opinion of the authors. 

{Contents after \S~\ref{sec:sub:eerllr} are new to the best of the authors' knowledge}. The shortcoming of EER is only briefly mentioned, for example, in privacy-ZEBRA. We give a detailed explanation and demonstration. Furthermore, the relationship between the ranking-based metric and perfect secrecy and the limitation of the calibration method in Privacy-ZEBRA are new.
\footnote{A Jupyter notebook is available to reproduce the results used in this paper: \url{https://doi.org/10.5281/zenodo.21868401}.}

\section{Threat Model and Perfect Secrecy}
\label{sec:secrecy}
Before we dive into the evaluation metrics, we need to agree on the information being protected as well as how an attacker plans to steal the information. This is referred to as the threat model, or attacker model~\cite{rahman_scenario_2024}. After explaining the threat model (\S~\ref{sec:sub:threat}), we go through the optimal Bayes decision that the attacker can make (\S~\ref{sec:sub:decision}), followed by perfect secrecy that gives the attacker no advantage in making the optimal decision (\S~\ref{sec:sub:perfect}). 

\subsection{Threat model}
\label{sec:sub:threat}
Intuitively, a defender wants to keep PII seclusive from unauthorized access by attackers (a.k.a \emph{privacy}), while not degrading other information flowing to downstream tasks (a.k.a. \emph{utility}). We focus on the evaluation of privacy protection and follow VoicePrivacy~\cite{tomashenko2020introducing} to define a threat model\textemdash a description of the interaction between the defender who builds a voice anonymization system and the attacker who tries to access the asset protected by the anonymization system. 
\begin{itemize}
    \item \textbf{Asset to be protected:} Speaker identity, i.e., who the speaker is in the input speech waveform. 
    \item \textbf{Attacker:} Adversary that attempts to gain information about protected speaker identity by simply guessing \emph{whether} the speaker identity in a speech waveform matches the speaker identity encoded in another speech waveform that the attacker has full access to. 
    The attacker may use an ASV or any binary classification system. The speech waveform with speaker identity to be protected is the \emph{probe}, and the waveform with an identity known to the attacker is the \emph{enrollment}.
    \item \textbf{Protection goal:} The defender builds a system to process the probe so that the attacker cannot gain information about the speaker identity in the processed probe. 
    This system is a (utterance-level) voice anonymization system.
    \item \textbf{Attacker knowledge:} 
    \begin{itemize}
        \item The full detail of the anonymization system built by the defender, except a secret that controls the behavior of the anonymization but does not depend on the probe.\footnote{For example, the secret corresponds to a random seed set by the defender in the selection-based anonymization system~\cite{srivastavaDesignChoicesXVector2020}.} 
        \item Access to a dataset of unprotected utterances (enrollment) from many speakers, covering the speaker being protected.
        \item Access to other training data and pretrained models.
    \end{itemize}
    \item \textbf{Attack success criterion:} The attacker correctly associates the speaker identity in the anonymized probe speech with the target speaker in the enrollment.
\end{itemize}
The last point is the topic of this paper. We do not include factors such as the attacker's knowledge (to become semi-informed~\cite{tomashenkoThird2026a}). 
Neither do we consider the evaluation of utility preservation~\cite{rahman_scenario_2024}. To simplify the discussion, we assume that the attacker uses an ASV system, but any binary classifier that discriminate speakers may be used. These details are not affecting or being affected by the following discussion. 

\subsection{Optimal decision by attacker}
\label{sec:sub:decision}
Given the above threat model, the attacker needs to guess or  make a decision whether the probe and enrollment $\evidence =(\evidence^{(p)}, \evidence^{(r)}) \in\mathcal{X}$ match in terms of the speaker identity. The true answer is binary $\mathcal{Y}=\{\symbolhp{}_\classtar{}, \symbolhp{}_\classnon{}\}$, where
\begin{align*} 
\symbolhp{}_\classtar{}:  & \, \evidence_{p} \text{ matches the speaker identity in }  \evidence_{r}, \\
\symbolhp{}_\classnon{}: & \,\evidence_{p} \text{does NOT match the speaker identity in }\evidence_{r}.
\end{align*} 
Attacker action $a$ can be from $\mathcal{A} = \{\acty, \actn\}$:
\begin{align*}
\acty: &\, \text{attacker claims that } \evidence_{p}  \text{ matches } \evidence_r,\\
\actn: &\, \text{attacker claims that } \evidence_{p}  \text{ does NOT match } \evidence_r.
\end{align*}
Since the attacker uses an ASV system when making the decision, the $\evidence$ is consumed and converted into a score $\score\in\mathbb{R}$, where the value indicates how likely $\evidence_p$ matches $\evidence_r$. Intuitively, the attacker should take $a=\acty$ if $\score$ is high enough.  

The attacker may take an incorrect action: a matched case ($H_\classtar{}$) may be falsely denied (or \emph{miss}), while non-matched cases ($H_\classnon{}$) may be falsely affirmed (or \emph{false acceptance}). Incorrect actions can be assigned decision costs.
Given an action $a\in\mathcal{A}$ and the ground-truth label $y\in\mathcal{Y}$, we use $C(a, y)\geq0$ to denote the decision-cost function $c: \mathcal{A}\times{\mathcal{Y}}\rightarrow\mathbb{R}^{+}$. An example is given in Table~\ref{tab:app_cost}.

Following the practice of Bayes decision~\cite{dudaPattern2001}, the attacker's conditional risk $R(a | \score)$ when taking an action $a$ given the input $\score$ is computed as
\begin{equation}
R(a | \score) = \sum_{y\in\mathcal{Y}}C(a, y) P(y | \score), 
\end{equation}
where $P(y | \score)$ is the posterior probability of class $y$ when observing $\score$. 
By plugging in the possible values of $C(a, y)$ listed in Table~\ref{tab:app_cost}, we can compute the conditional risks when taking the two possible actions
\begin{align}
R(\acty | \score) &= \costfanon{} P(\symbolhp{}_\classnon{} | \score), \label{eq:app_cost1} \\ 
R(\actn | \score) &= \costmiss{} P(\symbolhp{}_\classtar{} | \score). \label{eq:app_cost2}
\end{align}
The Bayes decision rule tells the attacker to select the action $a$ with a smaller risk~\cite{dudaPattern2001}, i.e.,
\begin{equation}
{a} = 
\begin{cases}
\acty, \quad \text{if } R(\actn | \score) > R(\acty | \score) \\
\actn, \quad \text{if } R(\actn | \score)  \leq R(\acty | \score) \\
\end{cases}.
\end{equation}
\begin{table}[!t]
\caption{Values of decision cost $C(a, y)$. Subscript $_\text{fa}$ and $_\text{fr}$ denote false $\acty$ and $\actn$, respectively.}
\vspace{-5mm}
\begin{center}
\begin{tabular}{rcll}
\toprule
 && Action = $\acty$ & Action = $\actn$\\
\midrule
True class = $\classtar{}$  && 0 & $\costmiss$ \\
True class = $\classnon{}$ && $\costfanon$ & 0\\
\bottomrule
\end{tabular}
\end{center}
\label{tab:app_cost}
\end{table}%
With Eqs.(\ref{eq:app_cost1}) and (\ref{eq:app_cost2}), the policy above can be re-written as 
\begin{equation}
\costmiss{} P(\symbolhp{}_\classtar{} | \score) \underset{\actn}{\overset{\acty}{\gtrless}}  \costfanon{} P(\symbolhp{}_\classnon{} | \score).
\end{equation}
From the Bayes' formula, the attacker knows that $P(\symbolhp{}_\classtar{} | \score) = p(\score | \symbolhp{}_\classtar{}) \pi_{\classtar{}}/p(x)$ and $P(\symbolhp{}_\classnon{} | \score)  
= p(\score | \symbolhp{}_\classnon{} ) \pi_{\classnon{}}/p(x)$,  
where $\pi_{\classtar{}}\triangleq P(H_\classtar{}) = 1-\pi_{\classnon{}}$ is the prior probability. The attacker can then further rewrite the policy in a log form as
\begin{equation}
 \log\frac{p( \score | \symbolhp{}_\classtar{})}{p( \score | \symbolhp{}_\classnon{} )} + \log\frac{\pi_\classtar}{(1-\pi_\classtar)}\underset{\actn}{\overset{\acty}{\gtrless}}  \log\frac{\costfanon{}}{\costmiss{}}.
\label{eq:app_policy_gen}
\end{equation}

The first term on the left-hand side of Eq.(\ref{eq:app_policy_gen}) is the LLR of the observed $\score$, $\mathsf{LLR}(\score) \triangleq \log\frac{p( \score | \symbolhp{}_\classtar{})}{p( \score | \symbolhp{}_\classnon{} )}$. This is the only term that depends on the input speech data (via $\score$). The logit $\log\frac{\pi_\classtar}{(1-\pi_\classtar)}$ and the term for decision costs are set \emph{in prior} by the attacker. In other words, when the attacker's ASV system is launched, the prior and decision costs are configured by the attacker. Given an input $\evidence_p$, the attacker compares $\evidence_p$ with a suspicious enrollment $\evidence_r$ using the classifier. The classifier produces $\score$, based on which the attacker makes a decision following Eq.(\ref{eq:app_policy_gen}). 

This is the optimal decision policy for the attacker.
It is the key to the discussion of how much the attacker gains from the input data and to what extent the defender can prevent that gain. 

\subsection{Bring in perfect secrecy}
\label{sec:sub:perfect}
Based on the threat model, what a defender can do via anonymization is to alter $\evidence^{(p)}$. \emph{Neither the enrollment ($\evidence^{(r)}$), the prior $\pi_\classtar$, nor the decision costs $C$ can be affected by the defender}. 
Therefore, the best protection that the defender can do is to give \emph{no extra information} to the attacker who makes a decision after observing $\evidence^{(p)}$ and computing $\score$. This is achieved when the protected $\evidence^{(p)}$ from the defender yields $\mathsf{LLR}(\score)=0$ for the attacker. Then the attacker only makes a decision based on the attacker's asserted priors and decision costs:
\begin{equation}
\underbrace{0}_{\text{Realm of defender, }\mathsf{LLR}(s)} + \underbrace{\log\frac{\pi_\classtar}{(1-\pi_\classtar)}\underset{\actn}{\overset{\acty}{\gtrless}}  \log\frac{\costfanon{}}{\costmiss{}}}_{\text{Realm of attacker}}.
\label{eq:app_policy_gen_perfect}
\end{equation}
To recap, the defender cannot step into the realm of the attacker, i.e., the prior or decision costs. The best defense is to give no extra information to the attacker, i.e., $\mathsf{LLR(s)}=0$. 

The above concept can also be explained using the Bayes formula. Making LLR equal to 0 means that the attacker has no extra information to convert the attacker's priors into the logit of posterior probabilities, which can be written as 
\begin{equation}
    \log\frac{P(\symbolhp{}_\classtar{} | \score)}{P(\symbolhp{}_\classnon{} | \score)} = \underbrace{\log\frac{p(\score | \symbolhp{}_\classtar{})}{p(\score | \symbolhp{}_\classnon{} )}}_\text{Realm of defender: make it 0} + \underbrace{\log\frac{\pi_\classtar{}}{1-\pi_{\classtar{}}}}_\text{Realm of attacker}.
\end{equation}
This is known as the perfect secrecy~\cite[\S~2.3]{stinson_cryptography_2005}\cite{shannon_communication_1949}.\footnote{In cryptography literature, it is written as $P(H_\classtar | \score)=P(H_\classtar)$ rather than in the form of logits and LLR~\cite[\S~2.3]{stinson_cryptography_2005}. }

\section{In Defense of Privacy-ZEBRA}
\label{sec:zebra}

Perfect secrecy for the defender is to drive every LLR to 0. An evaluation metric, however, takes the attacker's view, and the question is how close the LLR is to 0 for each probe utterance. 
With this in mind, we recap Privacy-ZEBRA in this section and then discuss EER and other metrics in the next sections.

\subsection{Measuring worst-case information leakage}
Privacy-ZEBRA~\cite{Nautsch2020} is the first framework adopting perfect secrecy for the voice anonymization task. 
A natural choice of measuring the information gained by the attacker is to compute $\epsilon(s) \triangleq \Big| \mathsf{LLR}(\score) \Big|$. 
On a protected test dataset $\mathcal{D}_\text{test}$, the maximum information leaked to the attacker, who uses an ASV system and an enrollment data set $\mathcal{D}_\text{enr}$, can be computed as
\begin{equation}
    \epsilon_\text{max} = {\max}_{s\in\mathcal{S}}\Big| \mathsf{LLR}(\score) \Big|,
    \label{eq:epsilonmax}
\end{equation}
where $\mathcal{S}\triangleq\{s =\text{ASV}(\evidence^{(p)}\evidence^{(r)}) | \evidence^{(p)}\in\mathcal{D}_\text{test}, \evidence^{(r)}\in\mathcal{D}_\text{enr}\}$ is the set of ASV scores. 

In practice, the evaluation recipe can be
\begin{enumerate}
    \item collecting scores $\score$ for the two classes ($H_\classtar$ and $H_\classnon$), for example, using the VoicePrivacy evaluation protocol and data,
    \item using a method to estimate $\mathsf{LLR}(\score)$ for each $\score$,
    \item and reporting $\epsilon_\text{max}$ on the whole evaluation set.\footnote{It is possible to compute $\epsilon_\text{max}$ over a subset of $\mathcal{D}_\text{test}$ that corresponds to a specific speaker. This gauges the privacy disclosure of that speaker.}
\end{enumerate}
Step 2 is necessary because $\mathsf{LLR}(\score)$ may not be equal to $\score$. 
Privacy-ZEBRA originally uses an algorithm called Pool Adjacent Violators (PAV)~\cite{brummer2013pav} to obtain the LLRs (\S~\ref{sec:calibration}). Another method is based on a similarity rank (\S~\ref{sec:rank}). 

This paper stands with privacy-ZEBRA for using $\epsilon_\text{max}$ as the evaluation metric of voice identity protection:
\begin{itemize}
    \item Perfect secrecy is independent of adversarial capabilities from a computation point of view. Even if the attacker has infinite computational power and time, $\epsilon_\text{max}=0$ means that the attacker gains no extra information after observing the anonymized data from the protected dataset. 
    \item The $\epsilon_\text{max}$ corresponds to the worst-case leakage. No information will be leaked more than $\epsilon_\text{max}$ for \emph{every} probe utterance in the protected evaluation dataset.
    \item The framework separates the role of prior knowledge of the attacker from the information depending on the defender (i.e., the LLRs in Eq.(\ref{eq:app_policy_gen_perfect})). No manual tuning of the decision threshold is needed to measure $\epsilon_\text{max}$. 
\end{itemize}

Note that Privacy-ZEBRA proposed an additional metric gauging expected privacy disclosure that is averaged over the attacker's asserted priors. Nevertheless, the perfect secrecy achieved by $\epsilon_\text{max}=0$ is the optimal solution for the expected privacy disclosure. 

\subsection{Privacy-ZEBRA and differential privacy}
\label{sec:sub:dp}
When talking about privacy, we cannot avoid the DP framework~\cite{dwork2014algorithmic}. 
Here, we tentatively suggest a link between $\epsilon_\text{max}$ defined by privacy-ZEBRA and the privacy loss defined by DP. 

In DP, the privacy loss incurred by observing $\score$~\cite[pp.18]{dwork2014algorithmic} is defined as $\frac{p(S=s | \mathcal{M}(\mathcal{D}_1))}{p(S=s| \mathcal{M}(\mathcal{D}_2))}$, where $S$ stands for a random variable of observation, $\score$ is the observed value, $\mathcal{M}$ is a protection mechanism, and $\mathcal{D}_1$ and $\mathcal{D}_2$ are `neighbor datasets'. The mechanism  $\mathcal{M}$ satisfies $\epsilon_\text{max}$-DP if the privacy loss for all possible values of $s$ is bounded by\footnote{Original notation writes $p(\mathcal{M}(\mathcal{D}_1) = s)$, which means observing $\score$ when $\mathcal{M}$ is applied to $\mathcal{D}_1$. To avoid confusion, we use $S$ and $s$ to denote the random variable and its observed value, respectively. Also note that DP literature usually defines the bound without the log and only includes the upper bound. The lower bound is implicitly included if we switch $\mathcal{D}_1$ and $\mathcal{D}_2$.}
\begin{equation}
    -\epsilon_\text{max} \leq \log \frac{P(S=s | \mathcal{M}(\mathcal{D}_1))}{P(S=s| \mathcal{M}(\mathcal{D}_2))} \leq \epsilon_\text{max}
    \label{eq:dp:loss}.
\end{equation}
The attacker cannot differentiate whether $s$ is from $\mathcal{D}_1$ or $\mathcal{D}_2$ with the difference in the probabilities larger than $\exp(\epsilon_\text{max})$. 

In the context of voice anonymization, the $\mathcal{M}$ corresponds to the voice anonymization system, and $\mathcal{D}_1$ and $\mathcal{D}_2$ can be interpreted as two versions of  $\mathcal{D}_\text{test}$: one containing the data from the target speaker ($H_\classtar$) and the other not ($H_\classnon$), or the reverse way.\footnote{In a well-known example of random YES-NO response for individual privacy protection~\cite{warnerRandomized1965}~\cite[Example 2]{karmitsa_comprehensive_2025}, the datasets can be interpreted as $\mathcal{D}_1 = \{\mathsf{YES}\}$ and $\mathcal{D}_2 = \{\mathsf{NO}\}$, and the privacy loss of observing YES, namely $p(\mathcal{M}(\{\mathsf{YES}\})=\text{YES})/p(\mathcal{M}(\{\mathsf{NO}\})=\text{YES})$, can be compute analytically.} 
If we agree with the definition of $\mathcal{D}_1$ and $\mathcal{D}_2$, we can see that a voice anonymization satisfies $\epsilon$-DP if $|\mathsf{LLR}(s)|$ is bounded by $\epsilon$, and 
anonymization reaching perfect secrecy is $0$-DP.

Studies in DP usually set $\epsilon_\text{max}>0$. The reason is that a DP mechanism has to keep some information, for example, when releasing medical records. The data to be released is processed as a whole.
For voice anonymization, however, $\epsilon_\text{max}=0$ is possible if the anonymization system perfectly disentangles the speaker-dependent and speaker-independent information. The speaker-dependent part then can be completely randomized, which ideally leads to $\epsilon_\text{max}=0$.\footnote{For applying DP to not only to the speaker embedding but also to content features, see the study in \cite{shamsabadi2022differentially}.}

\section{EER=50\% Is Not Perfect Secrecy}
\label{sec:eer}

\subsection{Highlighted messages}
We now discuss EER. Intuitively, an attacker who cannot link the speaker identity in anonymized utterance $\evidence^{(p)}$ with the attacker-known identity in enrollment $\evidence^{(r)}$ should obtain an EER equal to 50\%~\cite{tomashenko2020introducing}. 

In \S~\ref{sec:zebra}, we mentioned that the raw ASV scores and LLRs are not necessarily equal, which is decided by the goodness of score calibration~\cite{leeuwen13_interspeech}. Accordingly, the discussion on EER has to be separated for the cases wherein the attacker ASV system produces raw $\score$ and (ideally) calibrated $\mathsf{LLR}(s)$. 
The main messages are listed below.
\begin{itemize}
\item EER is measured over distributions of $\score$ or LLRs over the speakers. It does not measure an individual's privacy leakage. 
\item When EER is computed on ideally calibrated LLRs, EER=50\% is not well defined. 
\item When EER is computed on uncalibrated (or imperfectly calibrated) $\score$, EER=50\% is necessary but insufficient to reach $\epsilon_\text{max}=0$ for perfect secrecy. 
\end{itemize}
In short, \emph{EER=50\% does not necessarily mean perfect secrecy}. 
After describing the computation of EER and its properties, we present our arguments and demonstration using simulated data.

\subsection{Computation of EER}
\subsubsection{When attacker has ideal LLRs and their distributions}
Let $r\triangleq \mathsf{LLR}(s)\triangleq\log\frac{p( \score | \symbolhp{}_\classtar{})}{p( \score | \symbolhp{}_\classnon{} )}$ be a short-hand notation of the LLR value corresponding to $\score$. 
With the decision costs plugged in, the overall risk $R(\pi_\classtar, {\costfanon}, {\costmiss})$ when the attacker follows the optimal Bayes decision policy in Eq.(\ref{eq:app_policy_gen}) becomes:
\begin{align}
    &R(\pi_\classtar, {\costfanon}, {\costmiss}) 
    = \sum_{a\in\mathcal{A}}\int R(a | r) p(r)dr \\ 
    = &\int\sum_{a\in\mathcal{A}} \sum_{y\in\mathcal{Y}}C(a, y) P(y | r)p(r)dr\\
    =&\int_{\tau_\text{Bayes}}^{+\infty} {\costfanon}P(H_\classnon | r)p(r)d r + \int_{-\infty}^{\tau_\text{Bayes}} {\costmiss}P(H_\classtar | r)p(r)dr \\
    =&{\costfanon}(1-\pi_\classtar) \int_{\tau_\text{Bayes}}^{+\infty} p( r | H_\classnon) dr  +  {\costmiss}\pi_\classtar \int_{-\infty}^{\tau_\text{Bayes}}p(r |  H_\classtar )dr \label{eq:overallrisk_llr}\\
    =&{\costfanon}(1-\pi_\classtar) p_{\text{fa}}(\tau_\text{Bayes}) +  {\costmiss}\pi_\classtar p_{\text{miss}}(\tau_\text{Bayes}),
\end{align}
where $p_{\text{fa}}(\tau_\text{Bayes})$ and $p_{\text{miss}}(\tau_\text{Bayes})$ correspond to the probabilities of false $\acty$ and $\actn$, respectively,
\begin{align}
p_{\text{fa}}(\tau_\text{Bayes})&\triangleq\int_{\tau_\text{Bayes}}^{+\infty} p( r | H_\classnon) dr, \label{eq:pfa:llr}\\
p_\text{miss}(\tau_\text{Bayes})&\triangleq \int_{-\infty}^{\tau_\text{Bayes}}p(r |  H_\classtar )dr,\label{eq:pmiss:llr}
\end{align}
and where $\tau_\text{Bayes}=\log\Big({(1-\pi_\classtar)\costfanon{}}/({\pi_\classtar\costmiss{}})\Big)$ is the optimal Bayes decision threshold decided by the attacker's asserted prior and decision costs.\footnote{Row (\ref{eq:overallrisk_llr}) is obtained because  $P(H_\classnon | r)p(r)=(1-\pi_\classtar) p( r | H_\classnon)$ and $P(H_\classtar | r)p(r)=\pi_\classtar p( r | H_\classtar)$. Furthermore, the prior and decision costs are independent of $r$ and moved outside of the integration.}

The notion that the overall risk $R$ is a function of $\{\pi_\classtar, {\costfanon}, {\costmiss}\}$ reflects the fact that, given the $r$ and its distributions, the attacker's decision risk is only affected by these three attacker-asserted parameters. If the attacker alters them, the decision threshold $\tau_\text{Bayes}$ and the corresponding overall risk are changed automatically. 

It has been shown that~\cite[\S~2]{brummer2021out}, given $C_\text{fa}^\classnon={\costmiss}=1$, EER is the maximum overall risk when the attacker varies $\pi_\classtar$. 
\begin{align}
    \text{EER} &= \max_{\pi_\classtar} R(\pi_\classtar, {\costfanon}=1, {\costmiss}=1) \label{eq:eerdef}\\
    &= \max_{\pi_\classtar} \Big((1-\pi_\classtar) p_{\text{fa}}(\tau(\pi_\classtar)) +  \pi_\classtar p_{\text{miss}}(\tau(\pi_\classtar))\Big),
\end{align}
where the threshold $\tau(\pi_\classtar)=\log\frac{1-\pi_\classtar}{\pi_\classtar}$ is a function of $\pi_\classtar$. The $\pi_\classtar^*$ that yields the maximum corresponds to the threshold $\tau_\text{EER}=\log\frac{1-\pi_{\classtar}^*}{\pi_{\classtar}^*}$, which also satisfies $p_{\text{fa}}(\tau_\text{EER}) = p_\text{miss}(\tau_\text{EER})$,
hence the name equal error rate.

The above interpretation requires that the attacker uses LLRs, the threshold decided by the priors, and equal decision costs. 
The EER is the upper bound of the attacker's overall risk with respect to the attacker's asserted $\pi_\classtar$, \emph{not the LLRs or their distributions}. Therefore, the EER reflects the \emph{attacker's worst-case performance}. It is for this reason that privacy-ZEBRA argues that `the EER will hence reflect an unduly optimistic estimate of privacy protections'~\cite{Nautsch2020}. This is only one of the disadvantages of EER; others will be discussed later.

\subsubsection{When attacker has ideal LLRs but not their distributions}

In reality, the attacker does not know the true $p(r|H_\classtar)$ and $p(r|H_\classnon)$ and cannot compute the probabilities of false $\acty$ and $\actn$ by integration. The attacker has to estimate the error probabilities on a test set by counting. 

Let $\mathcal{D}_{r,\classtar}=\{r_i | H_\classtar\}$ and $\mathcal{D}_{r,\classnon}=\{r_i | H_\classnon\}$ be the LLRs of the $H_\classtar$ and $H_\classnon$ samples in a test set. 
Let a general threshold be $\tau$, which may be different from the optimal $\tau_\text{Bayes}$. The error rates are estimated by
\begin{equation}
    P_\text{fa}(\tau) = \frac{1}{|\mathcal{D}_{r,\classnon}|}\sum_{r_i\in\mathcal{D}_{r,\classnon}}\mathrm{1}(r_i>\tau),
    \label{eq:pfa_c:llr}
\end{equation}
\begin{equation}
    P_\text{miss}(\tau) = \frac{1}{|\mathcal{D}_{r,\classtar}|}\sum_{r_i\in\mathcal{D}_{r,\classtar}}\mathrm{1}(r_i<\tau),
    \label{eq:pmiss_c:llr}
\end{equation}
where $\mathrm{1}(\cdot)$ is an indicator function.
Plugging in the estimated error rates, the attacker gets the estimated overall risk
\begin{equation}
    \widehat{R}(\pi_\classtar, {\costfanon}, {\costmiss},\tau) = {\costfanon}(1-\pi_\classtar) P_{\text{fa}}(\tau) +  {\costmiss}\pi_\classtar P_{\text{miss}}(\tau).
\end{equation}
Note that this is the decision cost function (DCF), for example, used in NIST SRE~\cite[Eq.(2)]{nistNIST2016}.

Since $P_\text{fa}(\tau)$ and $P_\text{miss}(\tau)$ are not continuous functions of $\tau$, it may be infeasible to find a $\tau_\text{EER}$ so that $p_{\text{fa}}(\tau_\text{EER}) = p_\text{miss}(\tau_\text{EER})$. As an approximation, 
the threshold can be set to $\hat{\tau}_\text{EER} \approx \arg\min_{\tau} | P_\text{miss}(\tau) - P_\text{fa}(\tau) |$, and the EER can be
\begin{equation}
    \text{EER}\approx\Big(P_\text{miss}(\hat{\tau}_\text{EER}) + P_\text{fa}(\hat{\tau}_\text{EER})\Big)/2.
    \label{eq:eer:approximate}
\end{equation}

\subsubsection{When attacker has no ideal LLRs nor their distributions}
In this case, the EER is computed in the same way as the previous equation except that the $P_\text{fa}(\tau)$ and $P_\text{miss}(\tau)$ are counted based on the scores $\score$.
\footnote{If calibration is done via a `monotonic rising function', the EER value will not change~\cite[\S~3.1]{van2007introduction}. For example, a simple case is when the calibration function $\mathcal{C}(x)=ax+b$ with $a>0$, the threshold will be $C(\tau_\text{EER})$, and the EER stays the same.}


\subsection{EER=50\% vs. perfect secrecy}
\label{sec:sub:eerllr}
Now we discuss the relationship between EER=50\% and the perfect secrecy ($\epsilon_\text{max}=0$) or equivalently $r=0$ for all the test data. We also separate the cases depending on whether ideal LLRs are available. In each case, we consider whether EER=50\% yields perfect secrecy and vice versa.

\subsubsection{When attacker has ideal LLRs and their distributions}
\label{sec:sub:sub:eerllr}
Given $r=0$, EER is not computationally feasible.  
Since $r=0$ for all the data, both $p( r | \symbolhp{}_\classtar{})$ and $p( r | \symbolhp{}_\classnon{} )$ are the same Dirac delta function $\delta(r=0)$. 
Regardless of the decision threshold $\tau$,  
the attacker takes either $\acty$ or $\actn$ for all the test data. In such a case, the attacker obtains either $\{p_{\text{fa}}(\tau)=1, p_{\text{miss}}(\tau)=0\}$ or $\{p_{\text{fa}}(\tau)=0, p_{\text{miss}}(\tau)=1\}$.
The overall risk only takes two values,
\begin{equation}
    R(\pi_\classtar, {\costfanon}=1, {\costmiss}=1) \in \{(1-\pi_\classtar),  \pi_\classtar\},
\end{equation}
and the attacker \emph{cannot} find a $\tau_\text{EER}$ so that $p_{\text{miss}}(\tau_\text{EER})=p_{\text{miss}}(\tau_\text{EER})$. 
Furthermore, the definition in Eq.(\ref{eq:eerdef}) only yields $\max_{\pi_\classtar}R(\pi_\classtar, {\costfanon}=1, {\costmiss}=1)=1$.

On the other hand, EER=50\% is also ill-defined when $r$ is not fixed to be 0. 
According to the property that the LLR of LLR is LLR~\cite{leeuwen13_interspeech}, $r$ must obey $r=\log\frac{p( r | \symbolhp{}_\classtar{})}{p( r | \symbolhp{}_\classnon{} )}$ or,  equivalently,
\begin{equation}
    \exp(r)p( r | \symbolhp{}_\classnon{} ) = p( r | \symbolhp{}_\classtar{}).\label{eq:llr_transform}
\end{equation}
Suppose that we find $\tau_\text{EER}$ so that $p_{\text{fa}}(\tau_\text{EER}) = p_\text{miss}(\tau_\text{EER})=1/2$. 
Since $p_\text{miss}(\tau_\text{EER})=\int_{-\infty}^{\tau_{\text{EER}}} p( r | \symbolhp{}_\classtar{})dr$, by plugging in Eq.(\ref{eq:llr_transform}), the attack obtains
\begin{equation}
    \int_{-\infty}^{\tau_{\text{EER}}} \exp(r)p( r | \symbolhp{}_\classnon{} )dr = \int_{-\infty}^{\tau_{\text{EER}}} p( r | \symbolhp{}_\classtar{})dr =  \frac{1}{2}.
\label{eq:eer:llr1}
\end{equation}
If $\tau_{\text{EER}}\leq 0$, the attacker knows that $\exp(r)\leq 1, \forall{r}\in(-\infty, \tau_{\text{EER}}]$. Hence, the attacker gets an upper bound for the left side of the equation above: 
\begin{equation}
\int_{-\infty}^{\tau_{\text{EER}}} \exp(r)p( r | \symbolhp{}_\classnon{} )dr < \int_{-\infty}^{\tau_{\text{EER}}} p( r | \symbolhp{}_\classnon{} )dr = \frac{1}{2},
\label{eq:eer:llr2}
\end{equation}
where the right side is obtained because $\int_{-\infty}^{\tau_{\text{EER}}} p( r | \symbolhp{}_\classnon{} )dr = 1-\int_{\tau_{\text{EER}}}^{\infty} p( r | \symbolhp{}_\classnon{} )dr= 1-p_\text{fa}(\tau_\text{EER})=1/2$.
Now the attacker notices that Eqs.(\ref{eq:eer:llr1}) and (\ref{eq:eer:llr2}) are conflicting with each other. The conflict exists also for $\tau_{\text{EER}}> 0$. Hence, no $\tau_\text{EER}$ exists to reach EER=50\% when the attacker uses ideal LLRs that have non-zero values.

In short, the EER cannot be 50\% when the attacker have ideal LLRs. Measuring privacy protection performance using EERs estimated on ideal LLRs is not solid.

\subsubsection{When attacker has ideal LLRs but not their distributions}
In this case, the attacker has to estimate the error rates based on Eqs.(\ref{eq:pfa_c:llr}) and (\ref{eq:pmiss_c:llr}), but the conclusion is the same as the previous case. Given $r=0$ for all the test data, the attacker obtains either \{$P_\text{fa}=0, P_\text{miss}=1$\} or \{$P_\text{fa}=1, P_\text{miss}=0$\} and cannot reach an EER by definition. 
On the other hand, even if the attacker happens to find two sets of LLRs that produce EER=50\% based on Eqs.(\ref{eq:pfa_c:llr}-\ref{eq:eer:approximate}), if any LLR is not 0, the perfect secrecy is not reached by definition.

\subsubsection{When attacker uses raw scores and knows their true distributions}

Now let us consider the case when the attacker only uses $\score$ for making the decision, wherein $\score$ can be either uncalibrated or imperfectly calibrated. The conclusion is that perfect secrecy implies EER=50\% but not the other way around.

First, we show that perfect secrecy implies EER=50\%. 
By definition, perfect secrecy means that $r\triangleq\log\frac{p(s | H_\classtar)}{p(s|H_\classnon)}=0$, which implies that $p(s | H_\classtar) = p(s|H_\classnon)$ for any ${s}$. At a threshold $\tau_{\text{EER}}$ that leads to an equal value of false rejection and affirm, i.e., $\int_{-\infty}^{\tau_{\text{EER}}} p(\score \mid \symbolhp{}_\classtar{}) d\score =\int_{\tau_{\text{EER}}}^{+\infty} p(\score \mid \symbolhp{}_\classnon{}) d\score$, 
the attacker can replace $p(s | H_\classnon)$ with $ p(s|H_\classtar)$ and obtain
\begin{equation}
    \int_{-\infty}^{\tau_{\text{EER}}} p(\score \mid \symbolhp{}_\classtar{}) d\score = \int_{\tau_{\text{EER}}}^{+\infty} p(\score \mid \symbolhp{}_\classtar{}) d\score.
\end{equation}
Given $\int_{\tau_{\text{EER}}}^{+\infty}p(\score \mid \symbolhp{}_\classtar{}) d\score = 1-\int_{-\infty}^{\tau_{\text{EER}}} p(\score \mid \symbolhp{}_\classtar{})ds$, the attacker then obtains
\begin{equation}
    \int_{-\infty}^{\tau_{\text{EER}}} p(\score \mid \symbolhp{}_\classtar{}) d\score  = 1-\int_{-\infty}^{\tau_{\text{EER}}} p(\score \mid \symbolhp{}_\classtar{})ds.
\end{equation}
It is then obvious that $p_\text{miss}(\tau_\text{EER})=\int_{-\infty}^{\tau_{\text{EER}}} p(\score \mid \symbolhp{}_\classtar{})ds=1/2$. Similarly, it can be shown that $p_\text{fa}(\tau_\text{EER}) = 1/2$. Hence, EER is equal to 50\%. 

Note that the difference from the case in \S~\ref{sec:sub:sub:eerllr} is that $r=0$ does not require $s=0$. As long as the LLR reveals no information by $\log\frac{p( s | \symbolhp{}_\classtar{})}{p( s | \symbolhp{}_\classnon{} )}=0$, the value of $s$ can be non-zero. The $p(s|H_\classtar)$ and $p(s|H_\classnon)$ just need to be identical, and they can be any valid probability density function.

Now we show that EER=50\% does not guarantee perfect secrecy. This can be demonstrated using a counterexample where $p(\score \mid \symbolhp{}_\classtar{})$ and $p(\score \mid \symbolhp{}_\classnon{})$ are Gaussian distributions with the same mean value $\mu$ but different standard deviations $\sigma_\classtar{}\neq \sigma_\classnon{}$. 
It can be shown that $\tau_\text{EER} = \mu$ yields an EER of 50\%, however, the LLR is not always 0. Specifically, the LLR is equal to
\begin{equation}
    r\triangleq\log\frac{p(s | H_\classtar)}{p(s|H_\classnon)}=\log\frac{\sigma_\classnon}{\sigma_\classtar}+\frac{(s-\mu)^2}{2}(\frac{1}{\sigma_\classnon^2}-\frac{1}{\sigma_\classtar^2}).
    \label{eq:llr:example:gaussian}
\end{equation}
It is easy to see that $r\neq 0$ except for two specific values of $s$.

Figure~\ref{fig:eerllr} plots the counterexample where $\mu=0$, $\sigma_\classtar{} = 2$ and $\sigma_\classnon{} = 1$. There are 2,000 scores sampled from each Gaussian distribution. The oracle LLR (solid profile in a grey color in the bottom panel) shows the LLRs computed using Eq.(\ref{eq:llr:example:gaussian}). It can be observed that the oracle LLR curve crosses the line of perfect secrecy (LLR=0, dashed profile in a grey color), but the magnitude of LLR becomes larger as $s$ becomes farther away from $\mu=0$. 

\subsubsection{When attacker just has raw scores}
Unlike the previous case, the attacker cannot compute the LLR using the true score distributions and has to estimate it using other score calibration methods. However, the conclusion remains the same. 
If the estimated $r$ is 0 for any score, it means that the empirical score distributions of $H_\classtar$ and $H_\classnon$ overlap, and EER is 50\%.\footnote{The EER may not be exactly 50\% since the empirical error rates $P_\text{fa}$ and $P_\text{miss}$ are step-wise functions.}

\begin{figure}[t!]
    \centering
    \includegraphics[width=\linewidth]{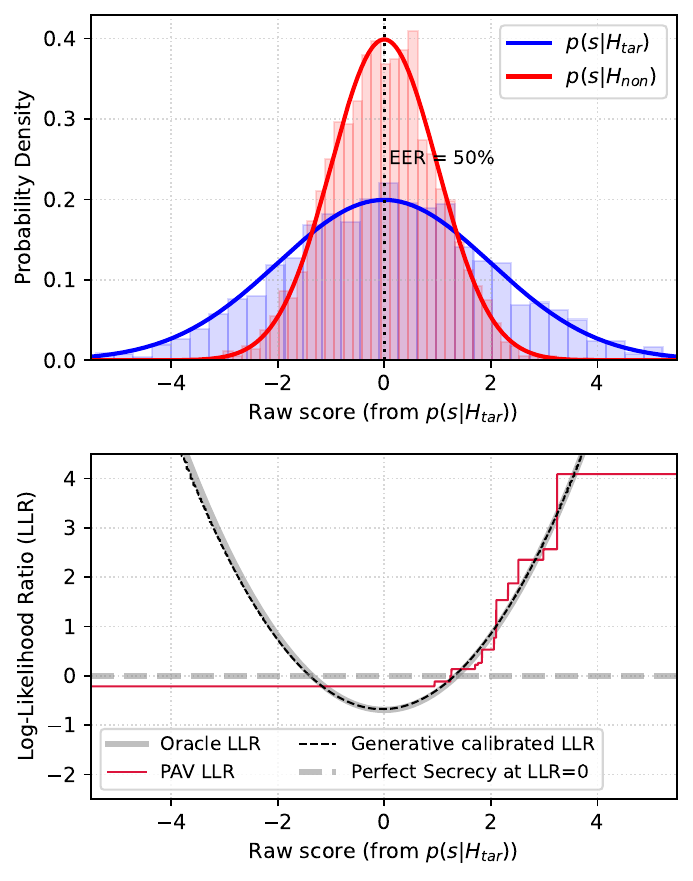}
    \vspace{-8mm}
    \caption{Demonstration of EER=50\% vs. perfect secrecy using simulated scores (top) and corresponding LLRs (bottom).}
    \label{fig:eerllr}
\end{figure}

On the other hand, EER=50\% does not mean $\mathsf{LLR}=0$. A counter example has been shown in Figure~\ref{fig:eerllr}. In this case, the attacker counts the histograms of the scores (in the top panel of Figure~\ref{fig:eerllr}) and estimates the LLRs using either a generative or non-linear calibration methods. In the former case, Gaussian distributions $p(s|H_\classtar)$ and $p(s|H_\classnon)$ are estimated from scores, and LLRs are computed by definition~\cite[\S~2.1]{VanLeeuwen2014}. The latter case uses the PAV algorithm~\cite{brummer2013pav}. 

The pros and cons of different LLR estimation methods are discussed in \S~\ref{sec:calibration}. Whichever the method is, the LLRs can be far from perfect secrecy (dashed grey line) even if the EER is 50\%.

\subsection{Summary on EER and related metrics}

In short, EER is a metric defined over score/LLR distributions. 
It considers the average case but not the worst-case information leakage.
In theory, if EER is computed given ideal LLRs, EER=50\% is not feasible. In reality, wherein LLRs are estimated or raw scores are used, even if the EER=50\% is reached, it does not guarantee perfect secrecy, as the counter-example in Figure~\ref{fig:eerllr} demonstrates. If EER=50\% does not necessarily imply perfect secrecy, the rationale for using EER is undermined.

The same argument can be mounted on other error-rate-based metrics, e.g., the GDPR-oriented Singling out probability~\cite{vauquierLegally2025} and VoxGuard~\cite{tsaprazlisVoxGuard2026a}.\footnote{It is interesting to note that VoxGuard~\cite{tsaprazlisVoxGuard2026a} is motivated by DP but does not use the LLR to audit the privacy loss (Eq.(\ref{eq:dp:loss})). Instead, it computes true/false positive rates at specific thresholds and use them to estimate the privacy loss.}\footnote{A linkability metric proposed in \cite{vauquierLegally2025} is threshold-free, but it still comes back to the empirical probability of finding a target speaker and hence can be counted as an error-rate-based metric.}
Note that different EER estimation methods (e.g., ROCCH-EER~\cite{brummerBOSARIS2013} rather than average in Eq.(\ref{eq:eer:approximate})) affect the accuracy of the estimated EER, but they do not change the fact that EER by definition does not measure the worst-case information leakage.



\begin{figure}[t!]
    \centering
    \includegraphics[width=\linewidth]{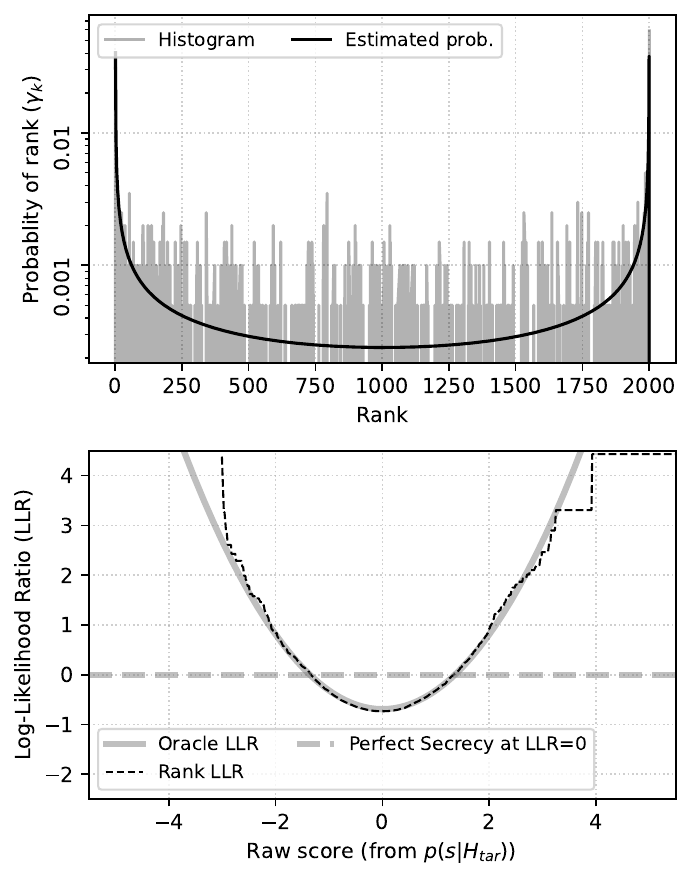}
    \vspace{-8mm}
    \caption{Illustration of rank of each score (top) and rank-based LLRs (bottom). Scores are from Figure~\ref{fig:eerllr}.}
    \label{fig:rankllr}
\end{figure}

\section{Similarity Rank is Compatible with Perfect Secrecy}
\label{sec:rank}
Recently, Bäckström et al. proposed a rank-based metric that measures PII disclosure~\cite{backstromPrivacy2026}. The core idea is to rank the speaker similarity scores and estimate the probability of a target-speaker score being at a particular rank. Then the probability can be used to measure the amount of leaked information. While it involves no LLR, we show that these rank probabilities can be cast into LLRs and linked to the concept of perfect secrecy. The optimal solution for perfect secrecy (zero LLR) is equivalent to the optimal case for the rank-based metric, wherein the probability of being at any rank is equal.

\subsection{How to compute rank and information leakage}
\label{sec:sub:rank}
Given ASV scores for both $H_\classtar$ and $H_\classnon$, we let the rank of the score of a matched case ($H_\classtar$) among the set of $N$ enrollment-trial pairs be $j \in [1, N]$. The probability of this rank $j$ is written as $\gamma_j = P_\text{R}(j|H_\classtar)$. Note that $\sum_{j=1}^{N}\gamma_j = 1$.
 
Given a set of scores, an attacker can compute the ranks, count the histogram, and estimate $\gamma_j, \forall j\in[1, N]$. Alternatively, the attacker can parameterize $\gamma_j$ via a Beta-binomial distribution and estimate the parameters of the distribution~\cite{backstromPrivacy2026}.
In either case, the information leakage when the anonymized target speaker is ranked $j$ on the dataset is measured by 
\begin{equation}
    \epsilon_j \triangleq -\log_2 N - \log_2\gamma_j
    \label{eq:rank:optimal}
\end{equation}
Ideally, if the rank of a target speaker's similarity score against all the enrollments is equally possible, i.e., $\gamma_j=\frac{1}{N}$, the information leakage becomes zero~\cite{backstromPrivacy2026}. 

\subsection{From rank probability to LLR}
While the rank-based $\epsilon_j$ involves no LLR, we now show that $\gamma_j$ can be used to compute the LLR and hence connected to the concept of perfect secrecy.

For a speaker with rank $k$ on the dataset, s/he can be either a matched target ($H_\classtar$) or non-target speaker ($H_\classnon$). 
The LLR of observing $k$ given the two hypotheses can be written as:
\begin{equation}
    \mathsf{LLR}(k) = \log\frac{P_\text{R}( k \mid H_\classtar)}{P_\text{R}( k \mid H_\classnon)}.
    \label{eq:rank:llr}
\end{equation}
By definition, $P_\text{R}( k\mid H_\classtar) \triangleq \gamma_k$. In contrast, the likelihood for $H_\classnon$ is derived in the following manner. Since a rank can only be occupied by exactly one speaker at a time, a specific non-target speaker can only occupy rank $k$ if the true target speaker resides at a different rank $r \neq k$. Let the event that the target speaker is ranked at $r$ be $\mathsf{T}_r$. Accordingly, the probability for rank $k$ to be a non-target speaker can be computed as 
\begin{align}
    P_\text{R}(k | H_\classnon) &= \sum_{r\neq k} P_\text{R}(k | \mathsf{T}_r) P_\text{R}(r | H_\classtar) = \sum_{r\neq k} P_\text{R}(k | \mathsf{T}_r) \gamma_r
\end{align}
If we assume that the remaining $N-1$ non-matched cases are equally possible, i.e., $P_\text{R}(k | \mathsf{T}_r)=\frac{1}{N-1}$, we obtain 
\begin{align}
    P_\text{R}(k | H_\classnon)  = \frac{1}{N-1}\sum_{r\neq k}  \gamma_r = \frac{1-\gamma_k}{N-1}.
    \label{eq:rank:non}
\end{align}
With Eqs.(\ref{eq:rank:non}) and (\ref{eq:rank:llr}), 
the LLR can now be computed as
\begin{equation}
    \mathsf{LLR}(k) = \log\frac{\gamma_k (N-1)}{1-\gamma_k}=\log(N - 1) + \log \left( \frac{\gamma_k}{1 - \gamma_k} \right)
    \label{eq:llrrank}.
\end{equation}
This is how the rank probability can be cast into LLR. 

The perfect secrecy requires that $\mathsf{LLR}(k)=0$ for all $k$. This is achieved if and only if $\gamma_k = \frac{1}{N}, \forall k\in[1, N]$, which is also the zero-leakage case according to Eq.(\ref{eq:rank:optimal}). Hence, the rank-based metric is connected to the concept of perfect secrecy.

From a practical point of view, the similarity rank can be used as a different way of estimating the LLRs. The evaluation recipe will be 1) follow the recipe in~\cite{backstromPrivacy2026} to estimate $\gamma_k$, 2) compute $\mathsf{LLR}(k)$ based on Eq.(\ref{eq:llrrank}), and 3) measure the worst-case information leakage by $\epsilon_\text{max}=\max_k|\mathsf{LLR}(k)|$. 
Based on the same set of scores plotted in Figure~\ref{fig:rankllr}, the rank histogram and estimated $\gamma_k$ are illustrated in the upper part of Figure~\ref{fig:rankllr}. The rank-LLRs are plotted in the bottom part of Figure~\ref{fig:rankllr}, which are close to the oracle LLR curve.  

\subsection{Summary on similarity rank}
In short, the similarity rank can be cast into LLRs and used to measure the information leakage. Zero information leakage from the similarity rank is equivalent to perfect secrecy.

\section{Notes on LLR Estimation Methods}
\label{sec:calibration}
As previous sections explain, in reality, the attacker has neither true LLRs nor true distributions of raw scores. The attacker has to calibrate the scores into LLR-like values.  

\subsection{Potential pitfalls when using PAV}
Privacy-ZEBRA uses the PAV algorithm~\cite{brummer2013pav} to calibrate scores into LLRs, after which the worst-case information leakage is computed using Eq.(\ref{eq:epsilonmax}). While PAV has been widely used for speaker verification~\cite{VanLeeuwen2014}, attention should be paid when it is applied for worst-case information leakage evaluation.

Given scores of $H_\classtar$ and $H_\classnon$, for example, from the evaluation set of the VoicePrivacy challenge, PAV first finds a monotonic step-wise mapping from the scores to the posterior probability $P(H_\classtar|s)$. 
Let $\mathcal{D}_{s,\classtar}=\{s_i |  H_\classtar\}$ and $\mathcal{D}_{s,\classnon}=\{s_i | H_\classnon\}$ be the scores of the $H_\classtar$ and $H_\classnon$ data in the test set. The PAV algorithm sorts all the scores and iteratively merges adjacent scores that violate a strictly increasing order. An example is shown at the top of Table~\ref{tab:pav}. Notice how the scores 0.30 from $H_\classtar$ and 0.40 from $H_\classnon$ violate the strictly increasing order, hence are pooled to a single `bin', and the resulting $P(H_\classtar|s)$ becomes 1/2. Given the PAV-estimated $P(H_\classtar|s)$, the LLR is computed by
\begin{equation}
    \mathsf{LLR}(s) = \log\frac{P(H_\classtar|s)}{1-P(H_\classtar|s)} - \log\frac{|\mathcal{D}_{s,\classtar}|}{|\mathcal{D}_{s,\classnon}|},
\end{equation}
where the second term is the logit of the empirical prior distribution, and $|\mathcal{D}|$ denotes the number of samples in the dataset.

The first pitfall is that a naive PAV implementation produces $P(H_\classtar|s)=1.0$ or 0.0 for the bins on the two ends, where the bins contain scores from either $H_\classtar$ or $H_\classnon$. Then, $\mathsf{LLR}(s)$ goes to $\pm\infty$, as the example 1 in Table~\ref{tab:pav} shows. 

Privacy-ZEBRA smooths the infinity values using \emph{Laplace succession rule}.\footnote{In implementation, after sorting the scores and getting the target $P(H_\classtar|s)$ (2nd row in Table~\ref{tab:pav}), two dummy target $P(H_\classtar|s)$ values [1, 0] are appended to both the left and right of the data array. After applying PAV, the dummy values are removed. See the example in the Jupyter notebook.} 
Let the right-most bin have $M$ scores from $H_\classtar$ only. Then the smoothed posterior probability becomes 
\begin{equation}
P_{\text{Lap}}(H_{\classtar} \mid s) = \frac{M + 1}{M + 2} < 1.0
\end{equation}
In Example 1 of Table~\ref{tab:pav}, the right-most bin has a single score from $\classtar$ (hence $M=1$). The smoothed $P_{\text{Lap}}(H_{\classtar} \mid s)=\frac{1+1}{1+2}=\frac{2}{3}$, and LLR becomes $\log 2=\log\frac{2/3}{1-2/3}$, where $\log\frac{|\mathcal{D}_{s,\classtar}|}{|\mathcal{D}_{s,\classnon}|}=0$.

The second pitfall, however, is caused by the smoothing. 
With the above two equations combined, we see that the estimated LLR for the largest or smallest raw score, which corresponds to the worst-case of information leakage, becomes
\begin{equation}
\mathsf{LLR}(s) = \log (M + 1)- \log\frac{|\mathcal{D}_{s,\classtar}|}{|\mathcal{D}_{s,\classnon}|}.
\end{equation}  
The LLR value is affected by the count of $M$ rather than the value of the largest or smallest score itself. Even if we increase the score $0.6$ in Example 1 to a very large value, the estimated LLR remains the same. 

Another interesting example is the Example 2 of Table~\ref{tab:pav}, which flips all the labels of Example 1. Although the scores of the two classes are as badly separated as in Example 1, the maximum LLR magnitude becomes 0.
In the Jupyter notebook, we further show that, if we replicate Examples 1 and 2 to longer sequences, the maximum magnitude of LLR remains $\log 2$ and $0$ for the two examples, respectively, even though the EERs become 50\% in both cases.  

In the bottom panel of Figure~\ref{fig:eerllr}, it is obvious that the PAV-estimated LLR has a step-wise shape, and its difference from the oracle LLR is huge on the two ends. 
In fact, the right-hand side is capped due to the smoothing\footnote{In the example, the right-most PAV pooled bin has around $M=55$ samples from $H_\classtar$. The LLR is capped at around $\log(55+1)\approx 4$.}, while the left-hand side is due to the pooled posterior probability that is close to 0.5. The left-hand side gives the illusion that the LLRs are close to 0. All the examples show how PAV can affect the estimation of the worst-case information leakage.

\begin{table}[!t]
\centering
\footnotesize
\setlength{\tabcolsep}{4pt}
\caption{Examples of PAV. Target $P(H_\classtar|s)$ is 1 for scores from $H_\classtar$, otherwise 0. For simplicity, equal priors are assumed so that the estimated LLR is $\log\frac{P(H_\classtar|s)}{1-P(H_\classtar|s)}$.}
\begin{tabular}{rrcccccc}
\toprule
\multirow{6}{*}{\rotatebox{90}{Example 1}}
& Sorted scores $s_i$
& 0.1&0.2&0.3&0.4&0.5&0.6 \\

& Target $P(H_\classtar|s)$
& 0&1&0&1&0&1\\

\cmidrule{2-8}

& PAVed $P(H_\classtar|s)$
& 0&$1/2$&${1}/{2}$&${1}/{2}$&${1}/{2}$&1\\

& Estimated LLR
& $-\infty$&$0$
&0&0&0&$+\infty$\\

\cmidrule{2-8}

& $P_{\rm Lap}(H_\classtar|s)$
&${1}/{3}$&${1}/{2}$
&${1}/{2}$&${1}/{2}$
&${2}/{3}$&${2}/{3}$\\

& Estimated LLR
&$-\log 2$&0
&0&0 &0&$\log 2$\\

\midrule

\multirow{4}{*}{\rotatebox{90}{Example 2}}

& Sorted scores $s_i$ 
& 0.1 &0.2 &0.3 &0.4 &0.5 &0.6 \\

& Target $P(H_\classtar|s)$
&1&0&1&0&1&0\\

\cmidrule{2-8}

& $P_{\rm Lap}(H_\classtar|s)$
&${1}/{2}$&${1}/{2}$&${1}/{2}$&${1}/{2}$&${1}/{2}$&${1}/{2}$\\

& Estimated LLR
& 0 & 0& 0 & 0& 0 & 0\\

\bottomrule
\end{tabular}
\label{tab:pav}
\end{table}

\subsection{Other score calibration methods}

In contrast to the non-parametric PAV algorithm, a generative approach assumes parametric score distributions $p(s|H_\classtar;\theta_\classtar)$ and $p(s|H_\classnon;\theta_\classnon)$. With the parameters $\{\theta_\classtar, \theta_\classnon\}$ estimated given the scores and labels in the test set, LLRs can be computed for each score as $\log \big(p(s|H_\classtar;\theta_\classtar)/p(s|H_\classnon;\theta_\classnon)\big)$. 

This generative method has been used in Figure~\ref{fig:eerllr} using Gaussian distributions. While the estimated LLRs are close to the oracle, the reason is that the simulated data is indeed from Gaussian distributions. 
In reality, we don't know the true distribution. Gaussian distributions may over-penalize the data on the two tails and affect $\epsilon_\text{max}$. In such a case, a heavy-tail t-distribution may better fit the data.

This is the dilemma when measuring perfect secrecy\textemdash the worst-case is decided by the extreme values of the LLR, but accurately estimating those LLR, which may be outliers, is challenging.  

\section{Demonstration Using VoicePrivacy Data}
\label{sec:vpc}
\begin{table}
\centering
\caption{EER and $\epsilon_\text{max}$ computed using scores from VPC 2024 systems. LLRs are estimated using either PAV, generative approach with Gaussian or t-distributions, or rank (\S~\ref{sec:rank}).}
\label{tab:vpc}
\begin{tabular}{rrrrrr}
\toprule
&   & \multicolumn{4}{c}{$\epsilon_\text{max}$ using estimated LLRs} \\
   \cmidrule{3-6}
& EER (\%) & PAV & Gen.(Gaus.) & Gen.(t.) & Rank \\
\midrule
B3 & 27.78 & 5.11 & 7.97 & 2.89 & 5.49 \\
B4 & 29.77 & 4.13 & 9.85 & 4.48 & 4.65 \\
B5 & 32.18 & 3.82 & 4.21 & 3.04 & 4.45 \\
T10-2 & 40.32 & 1.56 & 1.58 & 1.44 & 2.01 \\
T8-5 & 41.41 & 2.52 & 0.85 & 0.85 & 8.95 \\
T25-1 & 42.22 & 1.73 & 9.86 & 1.36 & 2.93 \\
T12-5 & 42.71 & 3.03 & 0.91 & 0.91 & 1.76 \\
\bottomrule
\end{tabular}
\end{table}
To compare EER and the maximum magnitude of LLRs on real data, 
we used the official scores released for the VoicePrivacy attacker challenge~\cite{tomashenko_first_2025}. Given the attacker ASV scores (in a semi-informed attacker setting), we pooled the female and male trials and computed EERs. We further used four methods to estimate the LLRs and computed $\epsilon_\text{max}$: PAV, the generative approach assuming Gaussian score distributions or t-distributions, and the rank-based approach. 

The results are plotted in Table~\ref{tab:vpc}. It is obvious that none of the systems achieved perfect secrecy. If we rank the systems based on $\epsilon_\text{max}$ (PAV), the order of the non-baseline systems will be slightly different, but all of them are better than the three baselines. This shows that the participants' submissions outperform the baselines from the point of view of worst-case leakage.

\begin{figure}[t!]
    \centering
    \includegraphics[width=\linewidth]{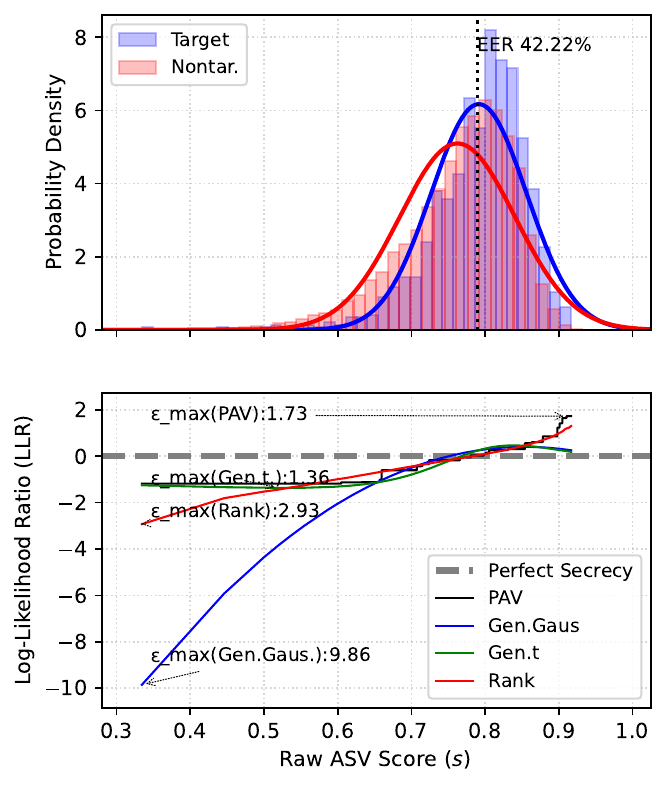}
    \vspace{-6mm}
    \caption{Raw scores of VPC system T25-1 (top) and corresponding LLRs (bottom).}
    \label{fig:vpcllr}
\end{figure}

The $\epsilon_\text{max}$ given LLRs estimated from the Gaussian-based generative approach seems to be occasionally abnormal. As Figure~\ref{fig:vpcllr} shows, T25-1's score distributions seem not to be well modeled by the Gaussians. In this case, using the long-tail t-distribution seems to obtain more reasonable LLRs. Note that since there is no ground-truth LLRs, we cannot tell which method estimates the LLRs the best. 
We at least see that a high EER\% does not necessarily mean perfect secrecy. 

\section{Conclusion}
\label{sec:conclude}
This paper revisits the privacy evaluation metric used for the voice anonymization task. Motivated by the fact that quite a few metrics have been proposed but not have been fully explained in terms of technical differences and underlying assumptions, this paper starts from the concept of perfect secrecy and shows how EER fails to capture the worst-case from a perfect secrecy point of view. This is further demonstrated using simulated data. The paper also links a newly proposed rank-metric to perfect secrecy and verifies the argument using simulated data. 

This study is in progress. Although this paper proposes one (potentially better) way of evaluating privacy protection performance, it does not address the issue of how to accurately estimate the LLRs or investigation of the sensitivity to various factors. This is the next step for future work.

\clearpage
\newpage

\section{Acknowledgment}
The authors thank the reviewer for the comments and suggestions. We have tried to address all the comments, but some of them cannot be reflected in the camera-ready version due to limited space. 
This work is partially supported by JST, PRESTO Grant (JPMJPR23P9), Japan. While the computation load was small, the experiment was done using TSUBAME4.0, Institute of Science Tokyo.

\section{Generative AI Use Disclosure}
Generative AI was used to check grammatical errors and prepare the code for figure plotting.

\bibliographystyle{ieeetr}
\bibliography{mybib}

\end{document}